\documentclass[conference]{IEEEtran}

\usepackage{cite}

\usepackage{multirow}

\ifCLASSINFOpdf
  \usepackage[pdftex]{graphicx}

\else

  \usepackage[dvips]{graphicx}

\fi

\usepackage[cmex10]{amsmath}

\usepackage{algorithm,algorithmic}

\usepackage[tight,footnotesize]{subfigure}

\usepackage{tikz}

\usepackage{url}

\usepackage{threeparttable}

\begin{document}

\title{A Memristor Crossbar-Based Computation Scheme with High Precision}

\author{\IEEEauthorblockN{}
\IEEEauthorblockA{ \\ Junyi Li Fudan University
\\ Fulin Peng Fudan University
  \\ Fan Yang Fudan University
  \\ Xuan Zeng  Fudan University
} }

\maketitle

\begin{abstract}
The memristor is promising to be the basic cell of next-generation computation systems.  Compared to the traditional MOSFET device, the memristor is efficient over energy and area.  But one of the biggest challenges faced with researchers is how to program a memristor's resistance precisely. Recently, an algorithm designed to save 8 valid bits in each memristor is proposed, but this is still not sufficient for precise computation.  In this paper, we propose a crossbar-based memristor computation scheme supporting precise computations whose operands have 32 valid bits.  As a brief introduction, in a multiplication with two operands, one operand is programmed as input signal, and the other operand is saved into a so-called crossbar structure, which contains a group of memristors, and each memristor saves several valid bits, usually one or two bits only. The computation results,i.e. the multiplication of the two operands, are contained in the outputs of the crossbar structure together with noise.  Analog-to-Digital Converters (ADCs) are then used to extract the valid bits, which are the most significant bits of outputs. These valid bits can be combined together with Digital-to-Analog Converters(DACs) to get the final results.  What's more, the precision of this computation scheme can be adjusted according to the definition of the user, 32 valid bits at most, so it is qualified for different  application contexts.
\end{abstract}

\IEEEpeerreviewmaketitle

\section{introduction}
HP Lab~\cite{DmitriB.Strukov2008} fabricated the first memristor in 2008, based on the predictions made by L.O.CHUA~\cite{O.CHUA1971} in 1971. The memristor made by the HP Lab is a TiO$_2$ thin-film structure. Many similar memristive devices and structures were tried in the next few years, including~\cite{ShimengYu;YiWu;YangChai;Provine2011},~\cite{ShimengYu2010},~\cite{Tsai2009},~\cite{Lee2008}, and~\cite{Dongsoo2006}.  Almost at the same time, the memristor show it potential over numerous application contexts.  For example, memristor-based non-volatile memory can achieve higher integration density than the traditional flash memory, and Memristor-CMOS hybrid structures are demonstrated to be efficient over reconfigurable computation.

Moreover, due to the similarity between the memristive and synaptic behaviors, memristor-based structures provide an efficient way to implement neuromorphic computation, especially the crossbar-based memristor structure is widely used, in this structure, memristors are allocated at the cross-points of the horizontal and vertical metal wires.  The memristors imitate the synaptic connections of the neural network models, and their resistances are programmed to hold the synaptic weights of the neural network models.In fact, the reason that these structures have good performance in neuromorphic computation is that neural network models are not sensitive to fluctuations of the memristor's resistance, which are cause by the large random variations in the writing process.  Recently, an algorithm has been proposed in~\cite{FabianAlibart2013},~\cite{MiaoHu2016} aiming to program memristor's resistance more precisely and achieve 8-bits writing precision at the cost of hundreds of times of writing and reading processes.  Although the writing precision has been significantly improved, it can still not satisfy the requirements of high-precision computation, where the errors should be at least less than $10^{-10}$.

In this paper, we propose a memristor crossbar-based computing scheme based on the result of~\cite{FabianAlibart2013}. Due to the limited precision of the memristor, we divide the multiple bits of computation into many groups. The computation of each group can achieve adequate accuracy by memristor crossbar-based structure with limited-precision of memristors.  Combining the computation results of all the groups together is challenging. Note that only several most significant bits of the results of a group are valid, therefore, we employ Analog-to-Digital Converters (ADCs) to extract these valid most significant bits. These valid bits are then combined together with DACs to obtain the final computation results with high precision. We tested our design with fixed-point multiplications. The experimental results demonstrate that our design can achieve $10^{-16}$ precision for 32-bit fixed-point multiplications.

The rest of the paper is organized as follows. In Section II, the background of memristor and its computation structure is reviewed. The detailed design will be presented from the aspects of theory and circuit structure in Section III and IV, respectively. The experimental results will be demonstrated in Section V. In Section VI, we conclude the paper.

\section{Background Review}
\subsection{Memristor}
\begin{figure}[tbp!]
\centering
\includegraphics [scale=0.7]{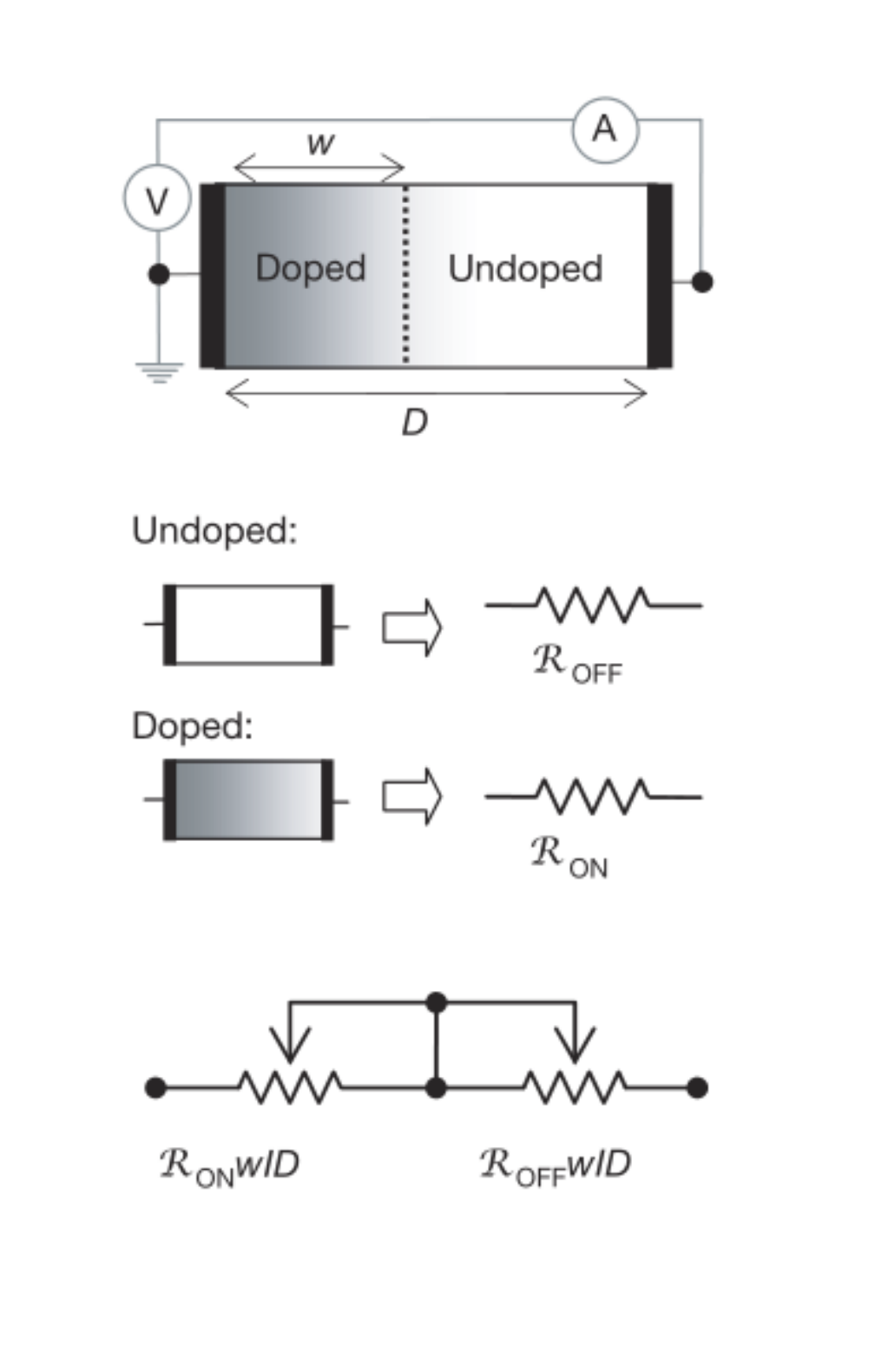}
\caption{The coupled variable-resistor model for a memristor.}
\label{fig:MemFigure}
\end{figure}

\begin{figure}[tbp!]
\centering
\includegraphics [scale=0.7]{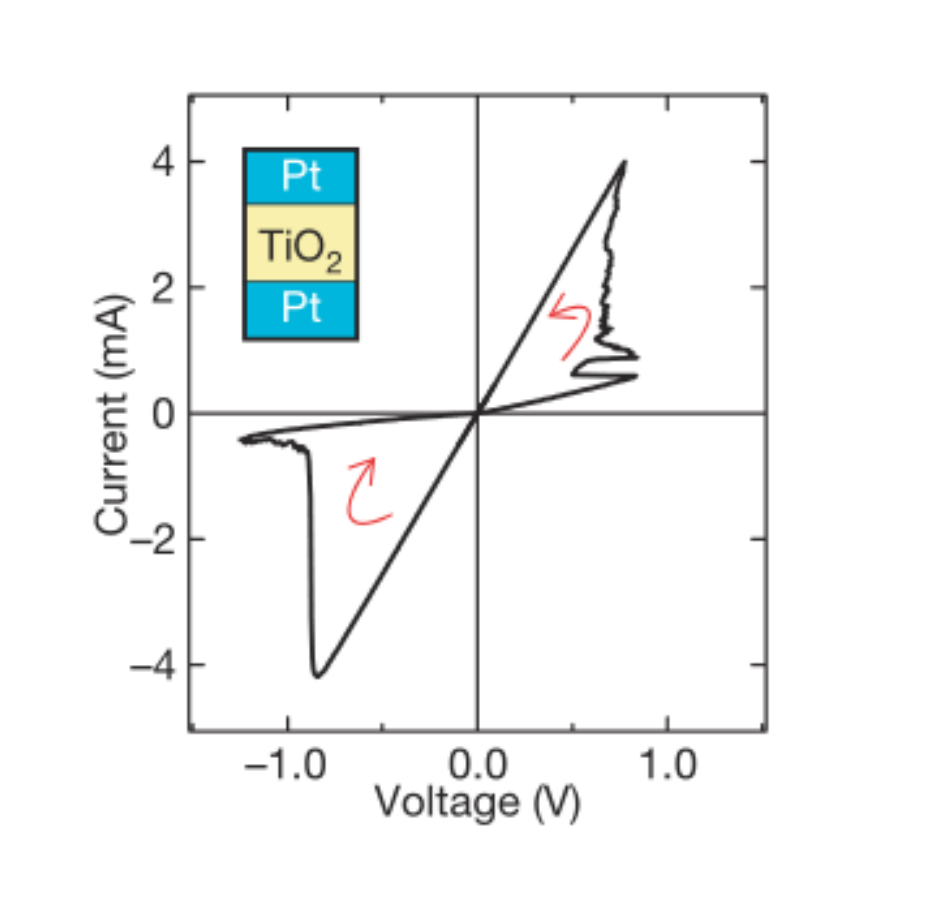}
\caption{I-V plot of a Pt-$TiO_2$-Pt device.}
\label{fig:Resistance}
\end{figure}

We will give a brief introduction to memristor firstly.  A typical structure of the memristor proposed by HP Lab in~\cite{DmitriB.Strukov2008} is shown in \ref{fig:MemFigure}. It is a semiconductor film, for example, copper oxide sandwiched between two metal contacts.  The resistance of the device is decided by two variable resistors connected in series, as shown in Figure\ref{fig:MemFigure}.  In fact, the oxide layer of the device can be divided into two regions, one is heavy-doped region, which is shown in figure in darker color, and the other is light-doped region, which is shown in lighter color.  Suppose the total length of the memristor is $D$, the length of the heavy-doped region is $w$ and the resistance of the this region is $R_{on}$, also, the resistance of light-doped region is $R_{off}$, besides, the mobility of the ion is $\mu_v$.  We have the following I-V relationship of the memristor.
\begin{equation} \label{iv_rel}
\begin{array}{rcl}
v(t) &=& \left(R_{on} \frac{w(t)}{D} + R_{off} \left(1-\frac{w(t)}{D}\right)\right) i(t), \\
\frac{dw(t)}{dt} &=& \mu_{v} \frac{R_{on}}{D} i(t),
\end{array}
\end{equation}
where $v(t)$ is the voltage applied on the memristor, and $i(t)$ is the current flowing through the memristor. Define the heavy-doped side of the memristor to be $p_1$ and the other side to be $p_2$. According to(\ref{iv_rel}), if the current flow from $p_1$ to $p_2$, the doped area will expand and $w$ become larger, which leads to the decrease of the resistance. In contrast, if the current flow from $p_2$ to $p_1$, the resistance of memristor will increase. The I-V relationship of a memristor is thus the curve shown in Figure \ref{fig:Resistance}~\cite{DmitriB.Strukov2008}.

\begin{figure}[tbp!]
\centering
\includegraphics[scale=0.4]{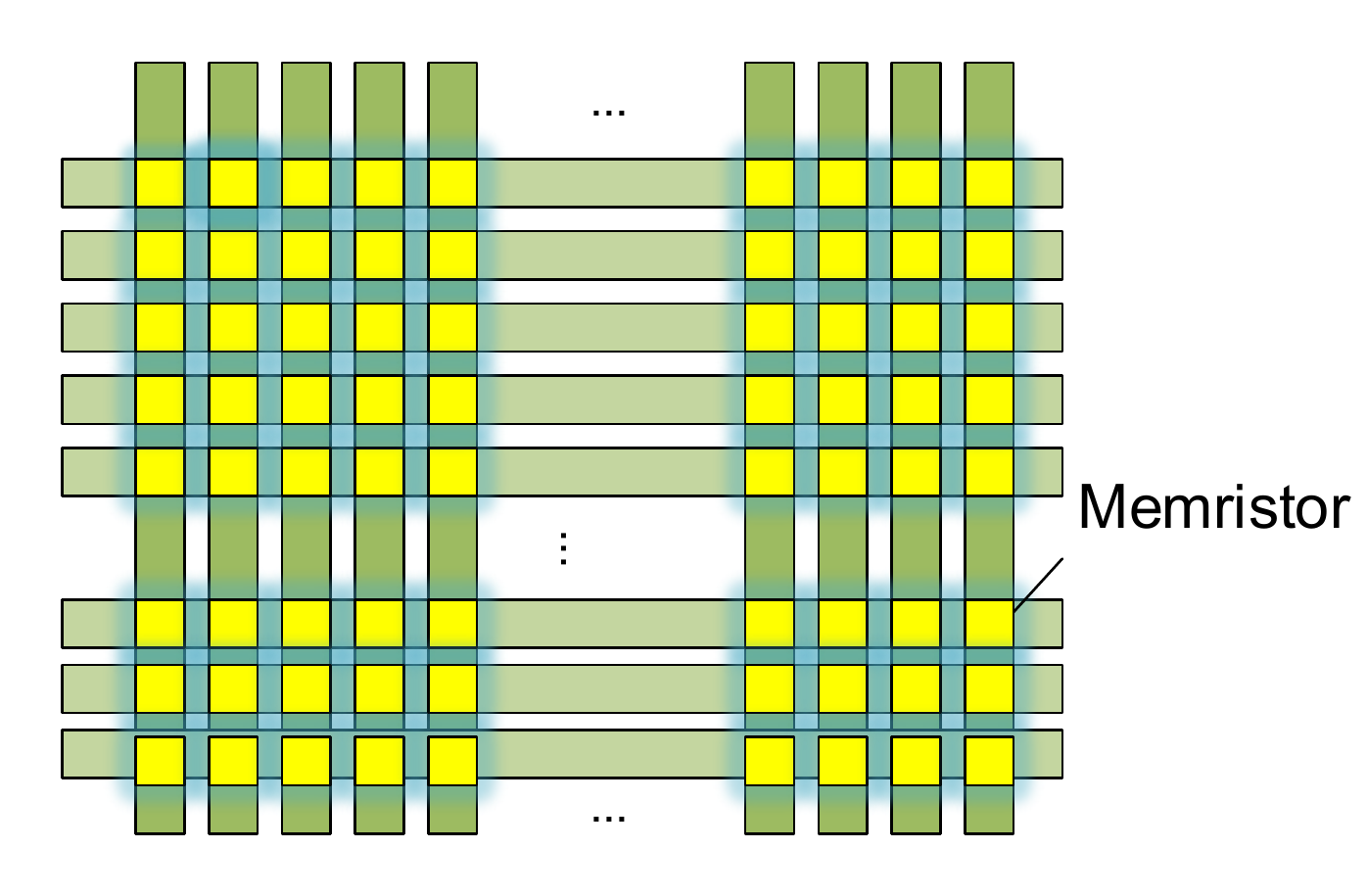}
\caption{normal crossbar array}
\label{fig:CrossBar}
\end{figure}

\subsection{Crossbar-based Structure for neuromorphic computation}
One of the most common applications of memristor is neuromorphic computation. The basic structure for neuromorphic computation is a crossbar array. Since the process of most artificial neural networks can be regarded as a series of matrix-vector mulplications, i.e.,
$$
\vec y=M \vec x,
$$
where $\vec y, \vec x$ are vectors, $M$ is a matrix. Such a computation can be accomplished by a memristor crossbar structure efficiently.  An example of the crossbar structure is shown in Figure \ref{fig:CrossBar}.

In the crossbar-based memristor structure, memristors are allocated at the cross-points of the horizontal and vertical metal wires. $\vec x$, $\vec y$ are the input and output of the crossbar array, respectively. The resistances (conductances) of the memristors can be viewed as the values of the elements of the matrix $M$. In neuromorphic computation, the resistances of memristors are programmed to be the weights of the synapses of artificial neural networks.  Based on this structure, a lot of artificial neural networks can be realized~\cite{MiaoHuRichardW2014}.

But we should note that the neuromorphic computation is not sensitive to the fluctuations of the synapses, which is caused by large random variations of the writing process of memristor.  As a result, the resistances of the memristors are not necessary to be in high precision.  But for other applications such as DSP algorithms, the resistances of the memristors must be programmed with much higher precision.  Recently, an algorithm has been proposed in~\cite{FabianAlibart2013} aiming to improve the precision of the memristors. It can achieve 8-bits writing precision with hundreds of times of writing and reading processes. Unfortunately, such a precision is still not sufficient for high-precision computation, where the error should be at least less than $10^{-10}$ in most cases.

\section{Data Representation}
In this section, we will present the theoretical analysis of our proposed crossbar-based memristor computing scheme.

\subsection{Representation of the High-precision Data by Multiple Memristors}
In prior works, a memristor usually corresponds to one element in matrix $M$. Aiming to improve the computation accuracy, we propose to store the high-precision data with a group of memristors. Our basic idea is that although the precision of one memristor is limited, the multiple memristors together can accurately represent data with high precisions. For the reason of convenience, we consider fixed-point unsigned numbers in our paper only. We also assume the target value $\alpha$ to be saved in memristors is within the range of $(0,1)$.
Firstly, we write $\alpha$ in binary form:
\begin{equation}
\alpha = \sum_{i=1}^{n} \alpha_i *2^{-i},
\end{equation}
where $\alpha_i \in \{0, 1\}$ is the $i$-th bit of the binary representation and the total number of bits is $n$. We aim to save $\alpha$ with $k$ memristors. For the reason of convenience, we assume that $n=k*m$. The value saved in the $j$-th memristor can be expressed as
\begin{equation} \label{dist_rep}
\beta_{j} = \sum_{l=1}^{m} \alpha_{(j-1)m+l} * 2^{-l} ~~~j = 1, \cdots, k,
\end{equation}
where $\beta_j$ is the data saved in the $j$-th memristor. So, $\alpha$ can be expressed by $\{\beta_1, \cdots, \beta_k\}$ as
\begin{equation} \label{dist_rep_all}
\alpha = \sum_{j=1}^{k} \beta_j * 2^{-(j-1)\times m}.
\end{equation}
It is important to point out that the least significant bit (LSB) of $\beta_{j} (1<j<k) $ is $2^{-m}$, in contrast to $2^{-km}$ in its original form.  In other words, the resistance of the memristor should have a precision of $2^{-km}$ to accurately save the original data $\alpha$. However, the data $\beta_j \{j = 1, \cdots, k\}$ can be accurately saved with only a precision of $2^{-m}$. At the cost of more memristors, the precision requirement is greatly reduced.

\subsection{Multiplication with Data Saved in Multiple Memresitors}
We consider following multiplication:
$$
z = x * y
$$
where $x, y, z$ are scalars represented by $n$ bits.  $x$ and $y$ can be written in the binary form as follows.
\begin{equation}
\begin{array}{rcl}
x &=&\displaystyle \sum_{i=1}^{n} x_i * 2^{-i},\\
y &=& \displaystyle \sum_{i=1}^{n} y_i * 2^{-i},
\end{array}
\end{equation}
where $x_i$ and $y_i$ are the $i$-th bit of $x$ and $y$, respectively.  Suppose $x$ is encoded by $k$ separated signals and $y$ is saved in $k$ memristors. They are expressed as follows.
\begin{equation} \label{y_exp}
\begin{array}{rcl}
x&=&\displaystyle \sum_{j=1}^{k} X_j * 2^{-(j-1) \times m},\\
y&=&\displaystyle  \sum_{j=1}^{k} Y_j * 2^{-(j-1)\times m},
\end{array}
\end{equation}
where:
\begin{equation}
X_j = \sum_{l=1}^{m} x_{jm+l} * 2^{-l},\\
Y_j = \sum_{l=1}^{m} y_{jm+l} * 2^{-l},
\end{equation}
If $x$ is encoded by the amplitudes of the sinusoidal wave,  $X_j$ will be the amplitude of the $j$-th wave.  While $Y_j$ is the value programmed in the $j$-th memristor.

Now, we consider the expressions of $z$ represented by $\{X_1, \cdots, X_k\}$ and $\{Y_1, \cdots, Y_k\}$. It can be easily verified that $z$ can be expressed as
\begin{equation}
z = \sum_{j=0}^{2k-2} Z_j * 2^{-j*m},
\end{equation}
where
\begin{equation} \label{z_exp}
Z_{j} = \sum_{p+q=j} X_p * Y_q ~~~j = 0, \cdots, 2k-2.
\end{equation}
Equation (\ref{z_exp}) indicates that we can accomplish the multiplication of $x$ and $y$ based on the sub-component expressions of $x$ and $y$. The rule is also very similar to the process of multiplying by hand.

\section{Circuit Implementation}
In this section, we will present the circuit implementation of our proposed approach.  An overview of the circuit structure is presented firstly, the implementations of the memristor crossbar and a so-called chain structures which is used to extract the valid bits from the outputs of the crossbar will be presented afterwards. We also assume that the number of valid bits of the data is $n$ and they are divided into $k$ groups for further processing, what's more, $n=k*m$.

\subsection{Overview}
\begin{figure}[tbp!]
\centering
\includegraphics[scale=0.5]{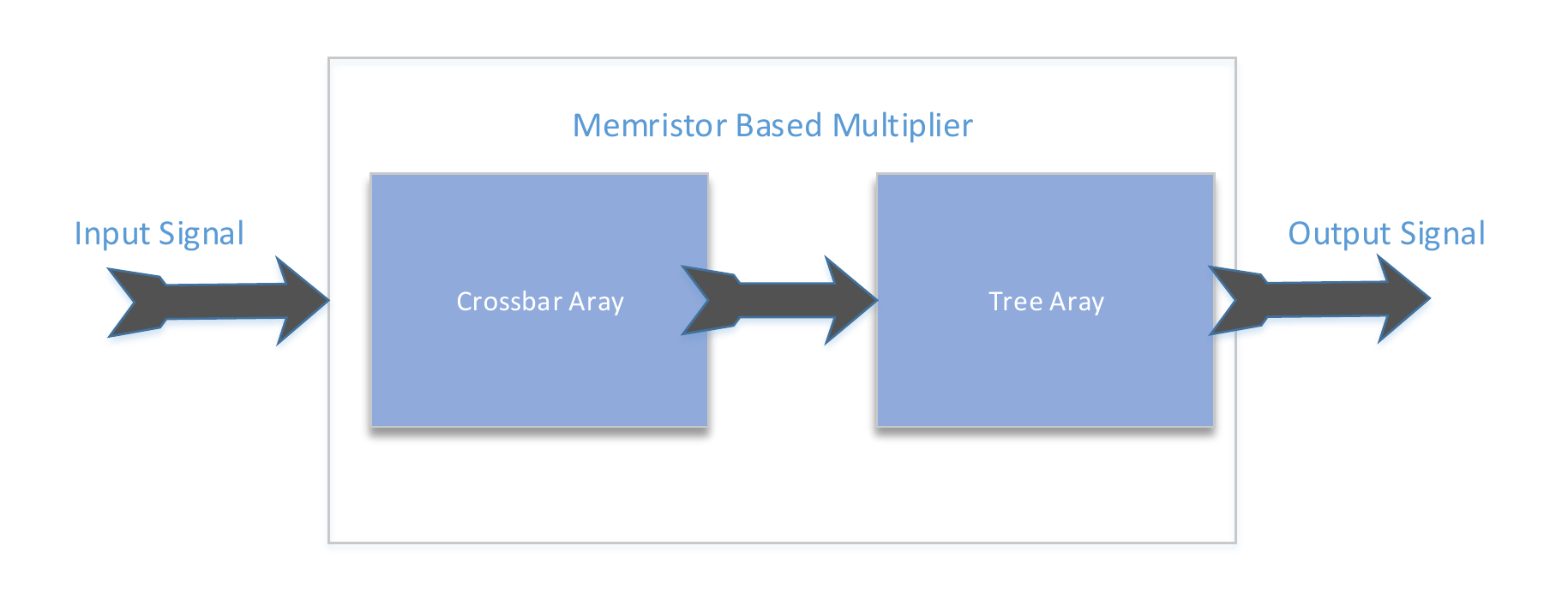}
\caption{Overview of our proposed memristor-based multiplier}
\label{fig:structure}
\end{figure}
The input of our structure is a scalar encoded by sinusoidal waves and the output is also a scalar.  As shown in Fig \ref{fig:structure}, our proposed structure consists of two main components. The first component is a crossbar array and the second component is a chain structure consisting of operational amplifier, ADCs and DACs.  The crossbar array is used to obtain $Z_j$ according to (\ref{z_exp}) and the chain structure is used to extract the valid bits of $Z_j$ obtained from the array structure, and combine these bits to get the final result.

\subsection{Memristor Crossbar Array}
\begin{figure}[tbp!]
\centering
\includegraphics[scale=0.4]{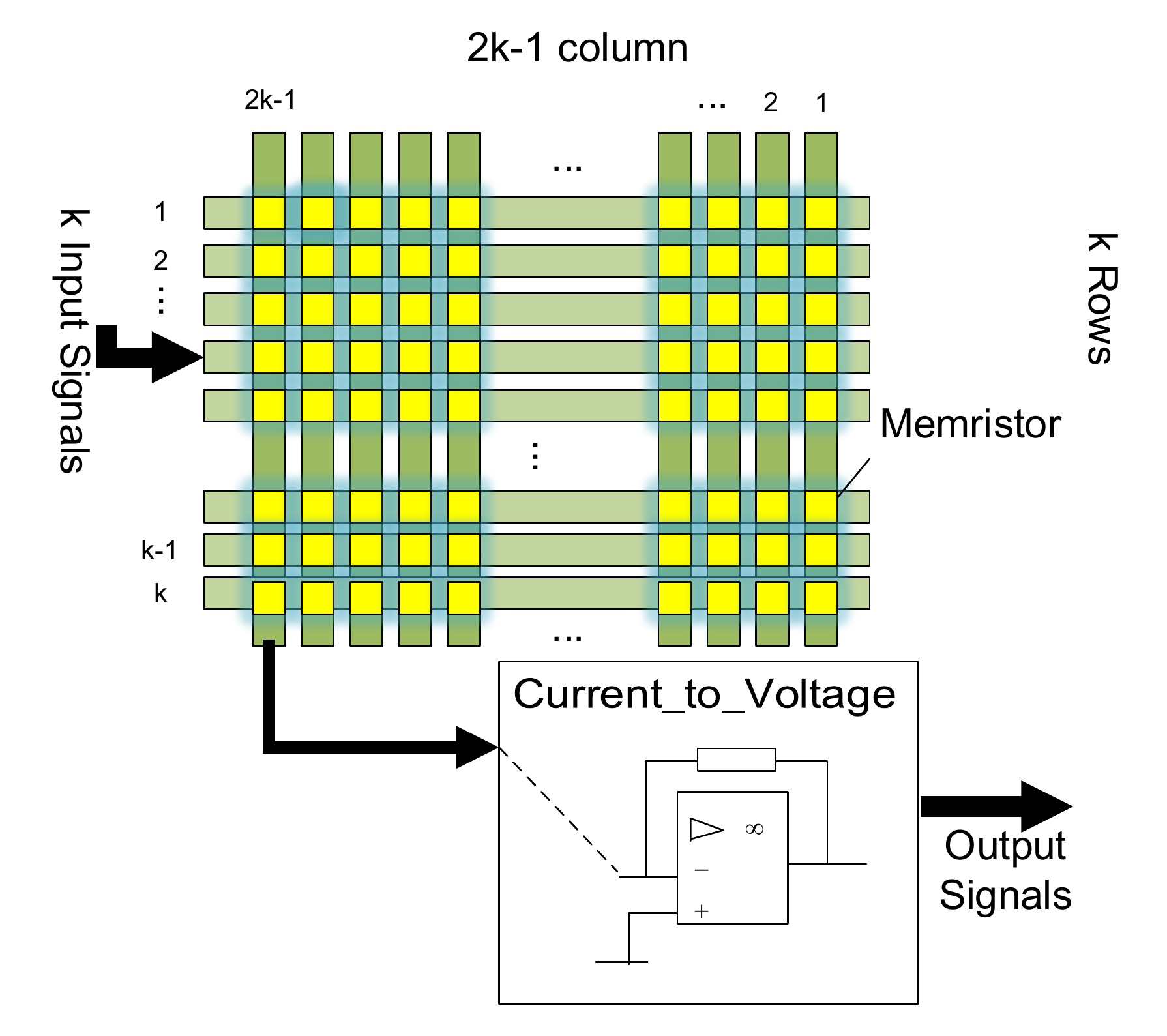}
\caption{Structure of Crossbar Array}
\label{fig:crossbar1}
\end{figure}
The structure of our crossbar array is similar to the traditional crossbar structure we mentioned before.  The input and output signals are all in the analog form, i.e. in the form of sinusoidal waves,  whose amplitude range from 0 to 1.  The amplitude of the sinusoidal waves encode the values of the input and output.  Memristors are allocated at the cross-points of the horizontal and vertical metal wires.  Instead of using a NMOS to cut off the sneak path~\cite{MiaoHuRichardW2014}, we use an operation-amplifier in each column to collect the current and transfer the current to the form of output voltage.  Thus, the input signals and output signals are all in the voltage form.  The structure of the crossbar structure is shown in Figure \ref{fig:crossbar1}.

Note that our crossbar is used to implement the computations of $Z_j$ as shown in (\ref{z_exp}).  Since $x$ and $y$ are divided into $k$ sub-components, the result of $z=xy$ thus consists of $2k-1$ sub-components.  As a result, our crossbar structure is a $(2k-1)\times k$ memristor array. The $k$ input signals are fed into the array from the left and represent the value of $x$ by their amplitudes.  More specifically, $X_i$ in (\ref{y_exp}) is encoded by the amplitude of the $i$-th sinusoidal waves and the $i$-th input signal is connected to the $i$-th row of the array as shown in Figure \ref{fig:crossbar1}.  The sub-components $\{Y_1, \cdots, Y_k\}$ are programmed in the memristors of the crossbar.  For the first row, the conductance of the 1st to $k$-th memristors are programmed with $\{Y_1, \cdots, Y_k\}$. Similarly, for the $k$-th row, the conductance of the $k$-th to $(2k-1)$-th memristors are programmed with $\{Y_1, \cdots, Y_k\}$. The conductance of all the rest memristors are set to be zeros.  In real applications, we can simply leave these cross-points open.

With such a crossbar structure, it can be easily verified that the current in the $i$-th column equals $Z_i$ as shown in (\ref{z_exp}).  An operation-amplifier is used in each column to transfer the current to the voltage output, However, two problems still need to be solved.  Firstly, the noises induced by the variations of the memristors remain in $Z_i$ and only the first several most significant bits are valid in $Z_i$. Secondly, the carries between the consecutive $Z_i$s are not tackled. These two issues will be addressed by a chain structure, which will be discussed in the next subsection.

\subsection{Chain Structure}
\begin{figure}[tbp!]
\centering
\includegraphics[scale=0.5]{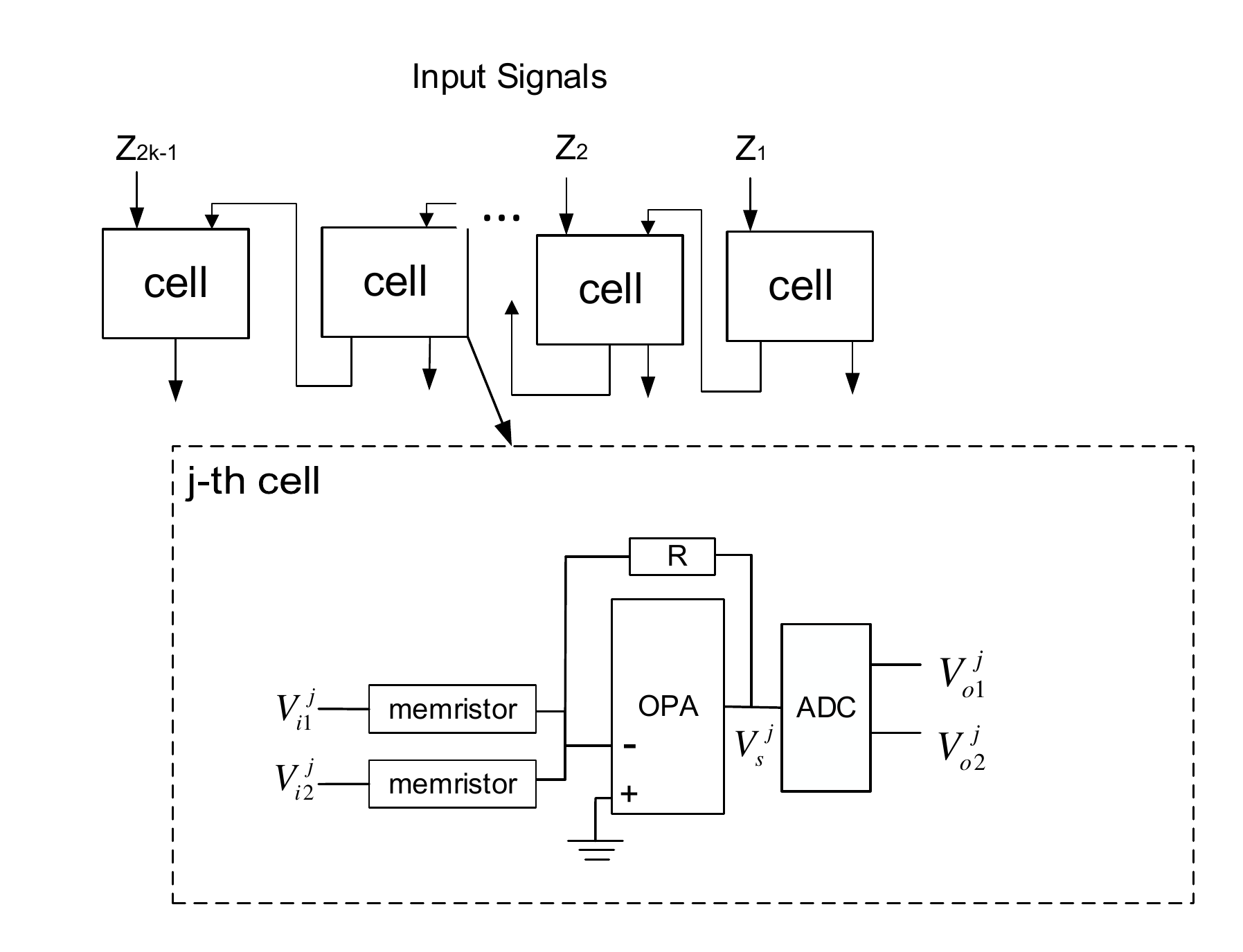}
\caption{The Chain Structure}
\label{fig:Tree}
\end{figure}

We use a chain structure to extract the valid most significant bits from $\{Z_1, \cdots, Z_{2k-1}\}$ and deal with the carry bits.  For the inputs $\{Z_1, \cdots, Z_{2k-1}\}$, each should have at least $p=m\rm{log}_2k$ effective bits.  The higher $p-m$ bits of each input are the carry bits, and the lowest $m$ bits are the output of the chain structure.  As shown in Figure \ref{fig:Tree}, the chain structure is similar to the traditional carry chains. The difference is that an ADC is used to extract the $p-m$ valid bits from each input, and the number of carry bits is not m but $p-m$. The $p-m$ carry bits are encoded by the amplitude of a voltage signal.   This signal is obtained directly from the inline DAC of the ADC.  The chain structure consists of $2k-1$ basic cells connected in series.  Each basic cell has two inputs and two outputs. Two inputs include the encoded $p-m$ carry bits from neighboring cell and $Z_i$ from the memristor crossbar array. Two outputs include the encoded $p-m$ carry bits, and $m$ bits in  binary form.

 If we assume that the two inputs of the $j$-th basic cell are $V_{i1}^j$ and $V_{i2}^j$, respectively. They are both in analog forms. Assume the outputs of the $j$-th basic cell are $V_{o1}^j$ and $V_{o2}^j$, respectively.  $V_{o1}^j$ in binary form is the output of the chain structure, and $V_{o2}^j$ in analog form containing the $p-m+1$ carry bits. According to the connections, we have the following relationship for the inputs and outputs
$$
\begin{array}{rcl}
V_{in2}^j &=& V_{o2}^{j-1}, \\
V_{in1}^j & = & Z_{j}.
\end{array}
$$

Because the $p-m$ carry bits are encoded by amplitude, $V_{in1}^j$ and $V_{in2}^j$ should be aligned and added together. Denote the sum of $V_{in1}^j$ and $V_{in2}^j$ by $V_{s}^j$, we have the following 

\begin{equation} 
\label{align}
V_{s}^j = V_{in1}^j + V_{in2}^j * 2^{-m},
\end{equation}
where $V_{in1}^j$ and $V_{in2}^j$ are aligned in the way of multiplying $V_{in2}^j$ by $2^{-m}$.  Equation (\ref{align}) can be implemented efficiently by a simple memristor structure as shown in Figure \ref{fig:Tree}. The two memristors are programmed with conductance 1 and $2^{-m}$, respectively.  The signal in current form from the memristor structure is then transformed to voltage form by a operational amplifier.  $V_{s}^j$ is then fed to an ADC to extract the effective bits.  After the first $p-m$ bits are extracted, it is transformed to its analog form by the inline DAC of ADC and taken as the output $V_{o2}^j$. The lowest m bits in binary form are the output $V_{o1}^j$ of the $i$-th cell.

\section{Experimental result}
We implement our proposed memristor-based high-precision structure. The memresitor model is based on~\cite{ChrisYakopcicPino2012}.  The CMOS circuits including operational amplifiers and ADCs.  HSPICE 2010 is used for circuit simulation.  In order to evaluate the impacts of the variations of the memristors and input signals, random variations are added to all the conductances of the memristors and input signals in our experiments.  We consider the fixed-point unsigned number in our experiments for convenience, and the range of the number is $(0, 1)$.

\subsection{An Illustrative Example}
In the first example, each memristor holds 2 valid bits. We set $x$ to be $0.8359375$, which can be accurately represented by 8 bits. Therefore, it is divided into 4 groups, and $\{X_j, j=1, \cdots, 4\}$ are listed as follows.
$$
\left(
  \begin{array}{cccc}
    0.75 & 0.25 & 0.25 & 0.5.  \\
  \end{array}
\right)
$$

$y$ is set to $0.42578125$ and it can be accurately represented by 8 bits too. $\{Y_j, j=1, \cdots, 4\}$ are listed as follows.
$$
\left(
  \begin{array}{cccc}
    0.25 & 0.5 & 0.75 & 0.25. \\
  \end{array}
\right)
$$

The result $z$ is expressed as
\begin{align}
  z &= xy \\
    &= 0.35592265137 \\
    &= 0.10110110001111_{(2)}.
\end{align}

In order to simulate the variations of the conductances of the memristors, Random noises are added to $\{Y_j, j=1, \cdots, 4\}$ as follows.
$$
\left(
  \begin{array}{ccccc}
    0.2546 & 0.2563 & 0.7510 & 0.2550  \\
  \end{array}
\right)
$$

In order to simulate the variations of the input signals, Random noises are added to $\{X_j, j=1, \cdots, 4\}$ as follows.
$$
\left(
  \begin{array}{ccccc}
    0.7509 & 0.2545 & 0.2564  & 0.5050  \\
  \end{array}
\right)
$$

The result of our multiplier is $0.10110110001111_{(2)}v$, which is equal to the standard result showed above.  This means that our structure is
able to make precise computation when the writing conductances of the memristors are not accurate and the input signals have noise.

\subsection{Multiplication of Scalars with Different Ranges}
We also consider the multiplication $z=xy$, where $x,y,z$ are all fixed-point numbers. $x,y$ have 16 effective bits. The ranges of $x$ and $y$ are set to (0,1). According to~\cite{FabianAlibart2013}, the writing conductance of a memristor can achieve 8-bit precision. We add random noises with absolute value less than $2^{-8}$ to the conductances of the memristors. In order to achieve $2^{-16}$ accuracy, a memristor is used to hold 1 effective bit in this example. In this case, $y$ with 16 effective bits are hold by 16 memristors.

Considering the outputs of the crossbar structure, 16 bits are added together in the worst case. Therefore, the accumulated error of random errors of 16 memristors with $2^{-8}$ precision would be smaller than $2^{-4}$ for $\{Z_1, \cdots, Z_16\}$, which means we can extract 4 effective bits from $\{Z_1, \cdots, Z_{16}\}$.  But 4 effective bits are enough to get the accurate result for the multiplication of scalars with 16 effective bits.

In order to test the precision of the multiplications, we random generate many combinations of $x$ and $y$ range from 0 to 1 and calculate the error compared with the accurate result.The testing data shows that we can realize the results with errors less than $2^{-16}$ at the rate of more than 99 percent and some of our testing data of $x$, $y$ and the error are shown in Table \ref{comparison}.
\begin{table}[tb] \centering
\caption{Multiplication of scalars with different ranges}
\label{comparison}
\begin{tabular}{|c|c|c|c|c|c|c|c|c|c|}
\hline $x$ &$y$ &error \\
\hline 0.033238043&0.135761887&0\\
\hline 0.033238043&0.265796779&0\\
\hline 0.033238043&0.391695303&0\\
\hline 0.033238043&0.504642826&0\\
\hline 0.033238043&0.630887185&0\\
\hline 0.033238043&0.81129003&0\\
\hline 0.033238043&0.97109508&0\\
\hline 0.135761887&0.265796779&6.94E-18\\
\hline 0.135761887&0.391695303&6.94E-18\\
\hline 0.135761887&0.504642826&0\\
\hline 0.135761887&0.630887185&0\\
\hline 0.135761887&0.81129003&0\\
\hline 0.135761887&0.97109508&2.78E-17\\
\hline 0.265796779&0.391695303&0\\
\hline 0.265796779&0.504642826&0\\
\hline 0.265796779&0.630887185&0\\
\hline 0.265796779&0.81129003&0\\
\hline 0.265796779&0.97109508&0\\
\hline 0.391695303&0.504642826&2.78E-17\\
\hline 0.391695303&0.630887185&2.78E-17\\
\hline 0.391695303&0.81129003&0\\
\hline 0.391695303&0.97109508&0\\
\hline 0.504642826&0.630887185&0\\
\hline 0.504642826&0.81129003&0\\
\hline 0.504642826&0.97109508&0\\
\hline 0.630887185&0.81129003&0\\
\hline 0.630887185&0.97109508&0\\
\hline 0.81129003&0.97109508&0\\
\hline
\end{tabular}
\end{table}
These results show that no matter the scalar is large or small, we can all conduct the desired precision. 
 %What's more, the proposed circuit can run at a the frequency up to 1 GHz, with an area of 6.68$um^2$ and the energy of 44.8mw

\subsection{Testing Results with Higher Precisions of the Memristors}
If the writing conductance of a memristor can achieve higher precision, e.g., 10-bit precision, we can accurately implement multiplication of $x$ and $y$ with 32 effective bits. In order to achieve $2^{-32}$ accuracy, a memristor is used to hold 1 effective bit in this case, and $y$ with 32 effective bits are hold by 32 memristors.

The accumulated error of random errors of 32 memristors with $2^{-10}$ precision would be smaller then $2^{-5}$ for $\{Z_1, \cdots, Z_32\}$, which means we can extract 5 effective bits from $\{Z_1, \cdots, Z_32\}$, while 5 effective bits are enough to represent the sum of 32 bits, which means we can get the accurate result for the multiplication of scalars with 32 effective bits.

In order to test the precision of the multiplications, we random generate many combinations of $x$ and $y$ range from 0 to 1 and calculate the error compared with the accurate result. Some of the testing data of $x$, $y$ and the error are shown in Table \ref{comparison2}. From the table, we can see that our proposed multiplier realizes the multiplication with the precision of $2^{-32}$.
\begin{table}[tb] \centering
\caption{Multiplication of scalars with different ranges}
\label{comparison2}
\begin{tabular}{|c|c|c|c|c|c|c|c|c|c|}
\hline $x$ &$y$ &error \\
\hline 0.033538067&0.134791476&0\\
\hline 0.033538067&0.266793781&0\\
\hline 0.033538067&0.371682311&0\\
\hline 0.033538067&0.512642927&0\\
\hline 0.033538067&0.670817231&0\\
\hline 0.033538067&0.83224021&0\\
\hline 0.033538067&0.97109508&0\\
\hline 0.135796238&0.266793781&6.78E-18\\
\hline 0.135796238&0.371682311&6.45E-18\\
\hline 0.135796238&0.512642927&0\\
\hline 0.135796238&0.670817231&0\\
\hline 0.135796238&0.83224021&0\\
\hline 0.135796238&0.92192513&2.36E-17\\
\hline 0.267796358&0.371682311&0\\
\hline 0.267796358&0.512642927&0\\
\hline 0.267796358&0.670817231&0\\
\hline 0.267796358&0.83224021&0\\
\hline 0.267796358&0.92192513&0\\
\hline 0.392696372&0.512642927&2.58E-17\\
\hline 0.392696372&0.670817231&2.77E-17\\
\hline 0.392696372&0.83224021&0\\
\hline 0.392696372&0.92192513&0\\
\hline 0.392696372&0.670817231&0\\
\hline 0.392696372&0.83224021&0\\
\hline 0.392696372&0.92192513&0\\
\hline 0.392696372&0.83224021&0\\
\hline 0.392696372&0.92192513&0\\
\hline 0.392696372&0.92192513&0\\
\hline
\end{tabular}
\end{table}

\section{Conclusion}
Compared to the traditional MOSFET devices, the memristor is much more efficient in energy and area, so it is considered as a promising candidate to overcome the Von Neumann bottleneck.  But to realize this dream, a memristor-based system must have the ability to conduct both precise computation and imprecise computation, otherwise it can only play the role as a GPU which means it can not replace the CPU and memory totally.  In this paper, we proposed a fundamental memristor-based computation scheme with the ability of precise computation.  We divide the multiple bits of computations into many groups. The computation of each group can achieve adequate accuracy.  Analog-to-Digital Converters (ADCs) are then used to extract these valid most significant bits. The valid bits are combined together to obtain the final computation result with high precision next.  Experimental results have demonstrated that if the conductances of memristors can achieve 8-bit precision as shown in~\cite{FabianAlibart2013}, our proposed approach can achieve accurate results for the multiplication of 16-bit fixed-point numbers. If the conductance of the memresitors can achieve 10-bit precision, our proposed approach can achieve accurate result for the multiplication of 32-bit fixed-point numbers. In the future work, we want to expand our structure to conduct float point computations.
\bibliographystyle{IEEEtran}
\bibliography{new}
\end{document}